# Selective Noise Suppression Methods Using Random SVPWM to Shape the Noise Spectrum of PMSMs

Jian Wen, Xiaobin Cheng, Peifeng Ji, Jun Yang, *Senior Member, IEEE,* and Feng Zhao, *Member, IEEE*

*Abstract*—Random pulse width modulation techniques are used in AC motors powered by two-level three-phase inverters, which cause a broadband spectrum of voltage, current, and electromagnetic force. The voltage distribution across a wide range of frequencies may increase the vibration and acoustic noise of motors. This study proposes two selective noise suppression (SNS) methods to eliminate voltage harmonics for specified frequencies. In the first method, the switching frequency is constant. The pulse position is calculated by the duty cycle of the current switching cycle. Both the pulse position and switching frequency are randomized in the second method. This involves creating a unique relationship among the switching frequency, pulse position, and duty cycle to shape the noise spectrum. Computer simulation and experimental results show that both methods effectively perform selective noise suppression at a specific frequency. The power spectrum density (PSD) using the second SNS method is more uniform near integer multiples of the switching frequency than that using random pulse width modulation techniques or the first SNS method. These methods provide a valuable reference for eliminating electromagnetic vibration and acoustic noises at resonant frequencies in motors.

## I. INTRODUCTION

NOISE problems in electronic control systems and motors have garnered increasing attention owing to the widespread application of motors in electric vehicles, the aerospace industry, and agricultural systems. Previous research shows that electromagnetic noise is the main constituent of most motors [1]. Although conventional space vector pulse width modulation (CSVPWM) is a widely used technique in AC motors powered by two-level three-phase inverters, it generates sideband tone noise in the motors. Random modulation techniques, such as random switching frequency SVPWM (RF-SVPWM) and random pulse position SVPWM (RP-SVPWM), have been proposed to spread the noise spectrum and effectively suppress the sideband noise [2]–[8]. The electromagnetic force has a broadband spectrum, which increases the possibility of motors resonating at the modal frequencies when these methods are used. Based on RF-SVPWM, previous research has proposed several methods for reducing voltage harmonics by establishing a relationship between the switching frequency and the position of voltage vector [9]–[12]. The implementation of these methods in motors can be complex and challenging, as the different duty cycle for each phase leads to different switching frequency for each phase. A selective current and electromagnetic force harmonic elimination method based on constant switching frequency was proposed in [13]. It is only effective under the low rotation speed conditions, and sideband noise at the switching frequency is significant.

This study proposes two selective noise suppression (SNS) methods at specified frequencies. The first method involves using random pulse position and fixed switching frequency, which theoretically has a lower minimum limit for the selection of the specified frequency and requires less memory occupation of the microcontroller unit (MCU) than the method proposed in [13]. In the second method, both the pulse position and switching frequency are randomized. A unique correlation is established among the switching frequency, pulse position, and duty cycle to shape the voltage spectrum. The second SNS method has two main advantages over previous methods. It spreads the voltage and current spectra more effectively, and is feasible for high modulation indices and rotation speeds. The proposed methods successfully achieved a gap in the noise spectrum.

The rest of this paper is structured as follows. In Section II, a voltage pulse waveform model using variable pulse position and switching frequency is developed in detail. Based on this model, two novel SNS methods are proposed to eliminate harmonic noise at specific frequencies. Computer simulation and experimental results are provided and analyzed in Sections III and IV to demonstrate the effectiveness of shaping the spectrum. Finally, the conclusions are discussed in Section V.

## II. UNDERLYING THEORY OF THE SELECTIVE NOISE SUPPRESSION METHOD TO SHAPE THE VOLTAGE SPECTRUM

The two-level three-phase voltage source inverter (VSI) circuit is shown in Fig. 1. Fig. 2 illustrates the switching pulse waveform of the inverter using SVPWM. The switching pulse in one cycle can be written as:

This work was supported by the Strategic Priority Research Program of Chinese Academy of Sciences under Grant XDC02020400. *(Corresponding author: Xiaobin Cheng.)*

Jian Wen, Xiaobin Cheng, Peifeng Ji, and Jun Yang are with the Key Laboratory of Noise and Vibration Research, Institute of Acoustics, Chinese Academy of Sciences, Beijing 100190, China, and also with the University of Chinese Academy of Sciences, Beijing 100049, China (e-mail: wenjian@mail.ioa.ac.cn; xb_cheng@mail.ioa.ac.cn; jipeifeng@mail.ioa.ac.cn; jyang@mail.ioa.ac.cn).

Feng Zhao is with the Institute of Electrical Engineering, Chinese Academy of Sciences, Beijing 100190, China (e-mail: zhaofeng@mail.iee.ac.cn).

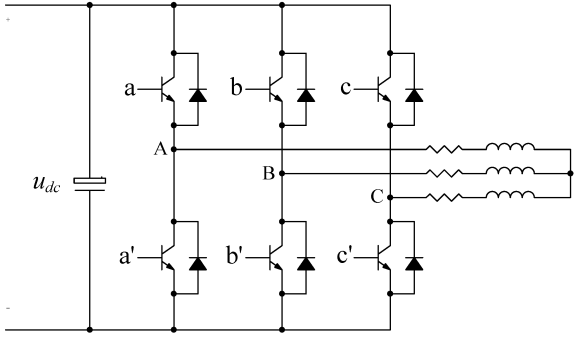

Fig. 1. Topology of the two-level three-phase VSI [4].

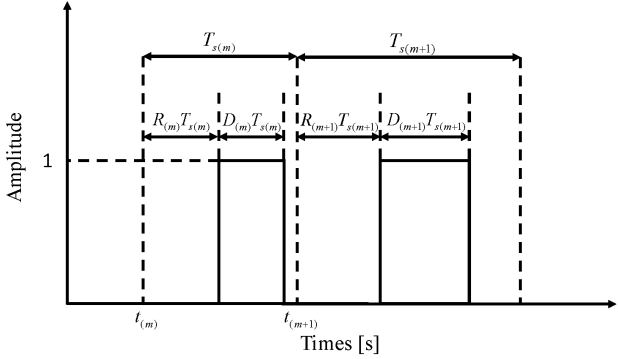

Fig. 2. Schematic diagram of the switching pulse waveform of an inverter for random pulse position and switching frequency.

$$x_m(t) = \begin{cases} 1, t_{(m)} + R_{(m)}T_{s(m)} < t < t_{(m)} + (R_{(m)} + D_{(m)})T_{s(m)} \\ 0, otherwise \end{cases} \quad (1)$$

Here, $T_{s(m)}$ is the $m^{th}$ switching cycle, $D_{(m)}$ and $R_{(m)}$ is the duty cycle and pulse position of the $m^{th}$ switching cycle, respectively. An amplitude of 1 represents the upper bridge arm being on and the lower bridge arm being off. Conversely, an amplitude of zero represents the upper bridge arm being off, with the lower bridge arm on.

Thus, the complete switching pulse train (taking phase A as an example) can be obtained:

$$x_A(t) = \lim_{N \to \infty} \sum_{m=1}^{N} x_{(m)}(t) \quad (2)$$

The switching pulse train $x_A(t)$ in the time domain can be transformed into the frequency domain by using Fourier transform:

$$\begin{aligned} X(f) &= \lim_{N \to \infty} \sum_{m=1}^{N} \int_{-\infty}^{+\infty} x_{(m)}(t)\exp(-j2\pi ft)dt \\ &= \lim_{N \to \infty} \sum_{m=1}^{N} \frac{\sin(\pi f D_{(m)} T_{s(m)})}{\pi f} \exp(-j2\pi f(t_{(m)} + R_{(m)} T_{s(m)} + \frac{D_{(m)}}{2} T_{s(m)})) \\ &= \frac{1}{2j\pi f} \lim_{N \to \infty} \sum_{m=1}^{N} \exp(-j2\pi f(t_{(m)} + R_{(m)} T_{s(m)})) \\ &\quad - \frac{1}{2j\pi f} \lim_{N \to \infty} \sum_{m=1}^{N} \exp(-j2\pi f(t_{(m)} + R_{(m)} T_{s(m)} + D_{(m)} T_{s(m)})) \end{aligned} \quad (3)$$

The exponential in (3) can be written as:

$$\begin{aligned} \alpha_{(m,f)} &= 2\pi f(t_{(m)} + R_{(m)} T_{s(m)} + D_{(m)} T_{s(m)}) \\ \beta_{(m,f)} &= 2\pi f(t_{(m)} + R_{(m)} T_{s(m)}) \end{aligned} \quad (4)$$

TABLE I
$\alpha_{(m,f_x)}$ AND $\beta_{(m,f_x)}$ FOR THREE POSITIONS

| Pulse Position | $\alpha_{(m,f_x)}$ | $\beta_{(m,f_x)}$ |
|---|---|---|
| $R_{(m)} = \frac{1-D_{(m)}}{2}$ | $2\pi f_x(t_{(m)} + \frac{1+D_{(m)}}{2}T_{s(m)})$ | $2\pi f_x(t_{(m)} + \frac{1-D_{(m)}}{2}T_{s(m)})$ |
| $R_{(m)} = 0$ | $2\pi f_x(t_{(m)} + D_{(m)}T_{s(m)})$ | $2\pi f_x t_{(m)}$ |
| $R_{(m)} = 1 - D_{(m)}$ | $2\pi f_x(t_{(m)} + T_{s(m)})$ | $2\pi f_x(t_{(m)} + (1-D_{(m)})T_{s(m)})$ |

TABLE II
THE EXPRESSION OF THE SWITCHING FREQUENCY FOR METHODS 1-3

| Method | Pulse Position | n=1 |
|---|---|---|
| $f_{s(m)} = \frac{D_{(m)}}{k} f_x$ | | |
| $f_{s(m+n)} = \frac{R_{(m+n)} + D_{(m+n)}}{\frac{k}{f_x} + \frac{R_{(m)}}{f_{s(m)}} - \sum_{i=0}^{n-1}\frac{1}{f_{s(m+i)}}}$ | $R_{(m)} = \frac{1-D_{(m)}}{2}$ | $f_{s(m+1)} = \frac{f_x(1+D_{(m+1)})}{2k - \frac{(1+D_{(m)})f_x}{f_{s(m)}}}$ |
| | $R_{(m)} = 0$ | $f_{s(m+1)} = \frac{f_x D_{(m+1)}}{k - \frac{f_x}{f_{s(m)}}}$ |
| | $R_{(m)} = 1 - D_{(m)}$ | $f_{s(m+1)} = \frac{f_x}{k - D_{(m)}\frac{f_x}{f_{s(m)}}}$ |
| $f_{s(m+n)} = \frac{R_{(m+n)}}{\frac{k}{f_x} + \frac{R_{(m)} + D_{(m)}}{f_{s(m)}} - \sum_{i=0}^{n-1}\frac{1}{f_{s(m+i)}}}$ | $R_{(m)} = \frac{1-D_{(m)}}{2}$ | $f_{s(m+1)} = \frac{f_x(1-D_{(m+1)})}{2k - \frac{(1-D_{(m)})f_x}{f_{s(m)}}}$ |
| | $R_{(m)} = 0$ | $f_{s(m+1)} = \frac{f_x}{k - \frac{(1-D_{(m)})f_x}{f_{s(m)}}}$ |
| | $R_{(m)} = 1 - D_{(m)}$ | $f_{s(m)} = \frac{1-D_{(m)}}{k} f_x$ |

$X(f)$ can be simplified by substituting (4) into (3):

$$X(f) = \frac{1}{2j\pi f}\left\{\lim_{N \to \infty}\sum_{m=1}^{N}\exp(-j\beta_{(m,f)}) - \lim_{N \to \infty}\sum_{m=1}^{N}\exp(-j\alpha_{(m,f)})\right\} \quad (5)$$

The correlations among the output phase voltages $u_A$, $u_B$, $u_C$, and the switching pulse train $x_A$, $x_B$, $x_C$ are as follows:

$$\begin{cases} u_A = \dfrac{u_{dc}(2x_A - x_B - x_C)}{3} \\ u_B = \dfrac{u_{dc}(2x_B - x_A - x_C)}{3} \\ u_C = \dfrac{u_{dc}(2x_C - x_A - x_B)}{3} \end{cases} \quad (6)$$

Here, $u_{dc}$ is the DC voltage link of the three-phase inverter.

A direct proportionality exists between the sound power and electromagnetic force, and their frequencies coincide [13]. Thus, if the switching pulse train $x_A$, $x_B$, $x_C$ spectra at a specific frequency $f_x$ are zero, then the voltage, electromagnetic force, and sound power can be eliminated at $f_x$. Let the value of (5) at a specific frequency $f = f_x$ be 0, then, the following condition needs to be satisfied:

$$X(f_x) = \frac{A}{2j\pi f_x}\left\{\lim_{N \to \infty}\sum_{m=1}^{N}\exp(-j\beta_{(m,f_x)}) - \lim_{N \to \infty}\sum_{m=1}^{N}\exp(-j\alpha_{(m,f_x)})\right\} = 0 \quad (7)$$

Based on whether the pulse position and switching frequency are random, there are three strategies to satisfy (7): the pulse position is fixed, while the switching frequency is random (based on RF-SVPWM); the pulse position is random, while the switching frequency is constant (based on RP-SVPWM); both the pulse position and switching frequency are random.

If the pulse position is fixed, there are typically three positions, namely, the front, center, and back [12]. Table I shows the values of $\alpha_{(m,f_x)}$ and $\beta_{(m,f_x)}$ at three pulse positions. Three methods are available to satisfy (7). Method 1 is $\alpha_{(m,f_x)} = \beta_{(m,f_x)} + 2k\pi$, where $k$ is an integer, and the $m^{th}$ term of $\sum \exp(-j\beta_{(m,f_x)})$ cancels out with the $m^{th}$ term of $\sum \exp(-j\alpha_{(m,f_x)})$. Method 2 is $\alpha_{(m+n,f_x)} = \beta_{(m,f_x)} + 2k\pi$, wherein the $m^{th}$ term of $\sum \exp(-j\beta_{(m,f_x)})$ cancels out with the $(m+n)^{th}$ term of $\sum \exp(-j\alpha_{(m,f_x)})$ [11], [12]. Method 3 is $\beta_{(m+n,f_x)} = \alpha_{(m,f_x)} + 2k\pi$, which indicates that the $(m+n)^{th}$ term of $\sum \exp(-j\beta_{(m,f_x)})$ cancels out with the $m^{th}$ term of $\sum \exp(-j\alpha_{(m,f_x)})$ [10], [12]. The correlation among the switching frequency, duty cycle, and pulse position can be established by substituting (4) into the aforementioned three methods, as shown in Table II, where $f_{s(m)} = 1/T_{s(m)}$.

*A. SNS method based on RP-SVPWM (SNS-RP-SVPWM)*

In this subsection, we derive a unique formula for eliminating the voltage harmonic at a specific frequency based on RP-SVPWM. Due to fixed switching frequency, $f_{s(m)}$ is constant and equal to $f_s$ ($T_{s(m)} = T_s$). Then, (4) is equivalent to:

$$\alpha_{(m,f)} = 2\pi f(t_{(m)} + R_{(m)}T_s + D_{(m)}T_s) \\ \beta_{(m,f)} = 2\pi f(t_{(m)} + R_{(m)}T_s) \quad (8)$$

Substituting (8) into $\alpha_{(m+n,f_x)} = \beta_{(m,f_x)} + 2k\pi$, we obtain:

$$2\pi f_x(t_{(m+n)} + R_{(m+n)}T_s + D_{(m+n)}T_s) = 2\pi f_x(t_{(m)} + R_{(m)}T_s) + 2k\pi \quad (9)$$

Thus,

$$R_{(m+n)} = \frac{k}{f_x}f_s + (R_{(m)} - D_{(m+n)} - n) \quad (10)$$

In this study, $n = 1$, (10) can be simplified to:

$$R_{(m+1)} = \frac{k}{f_x}f_s + (R_{(m)} - D_{(m+1)} - 1) \quad (11)$$

where $k$ is an integer.

The minimum and maximum limits of the pulse position $R_{(m+1)}$ are at the front and back of the $(m+1)^{th}$ switching cycle, respectively. Thus, it can be concluded that:

$$0 \leq R_{(m+1)} \leq 1 - D_{(m+1)} \quad (12)$$

The range of $k$ can be obtained by substituting (11) into (12):

$$\frac{f_x}{f_s}(1 - R_{(m)} + D_{(m+1)}) \leq k \leq \frac{f_x}{f_s}(2 - R_{(m)}) \quad (13)$$

Considering that $k$ is a positive integer, we obtain:

$$k_{\min} = \left\lceil \frac{f_x}{f_s}(1 - R_{(m)} + D_{(m+1)}) \right\rceil, \quad k_{\max} = \left\lfloor \frac{f_x}{f_s}(2 - R_{(m)}) \right\rfloor \quad (14)$$

where "$\lceil \ \rceil$" is the ceiling, and "$\lfloor \ \rfloor$" is the floor. Integer $k$ is selected from $K = \{k_{\min}, k_{\min}+1, \cdots, k_{\max}\}$ with equal probability.

(11) shows that the pulse position is determined by the duty cycle of the current switching cycle in SNS-RP-SVPWM. Only the formula for $\alpha_{(m+n,f_x)} = \beta_{(m,f_x)} + 2k\pi$ is derived, as the formula for $\beta_{(m+n,f_x)} = \alpha_{(m,f_x)} + 2k\pi$ was given in [13]:

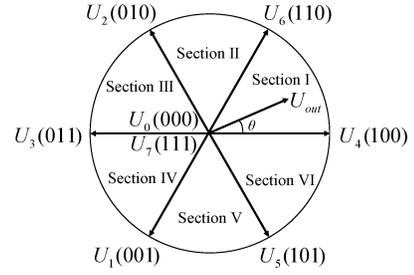

Fig. 3. SVPWM diagram.

TABLE III
DUTY CYCLE OF 5-SEGMENT SVPWM

| Section | $D_A$ | $D_B$ | $D_C$ |
|---|---|---|---|
| I | $\frac{T_6+T_7}{T_s} = M\sin(\theta+\frac{\pi}{3})$ | $\frac{T_6}{T_s} = M\sin(\theta)$ | 0 |
| II | $\frac{T_6}{T_s} = M\sin(\theta+\frac{\pi}{3})$ | $\frac{T_2+T_6}{T_s} = M\sin(\theta)$ | 0 |
| III | 0 | $\frac{T_2+T_3}{T_s} = M\sin(\theta-\frac{\pi}{3})$ | $\frac{T_3}{T_s} = -M\sin(\theta+\frac{\pi}{3})$ |
| IV | 0 | $\frac{T_3}{T_s} = M\sin(\theta-\frac{\pi}{3})$ | $\frac{T_1+T_3}{T_s} = -M\sin(\theta+\frac{\pi}{3})$ |
| V | $\frac{T_5}{T_s} = -M\sin(\theta-\frac{\pi}{3})$ | 0 | $\frac{T_1+T_5}{T_s} = -M\sin(\theta)$ |
| VI | $\frac{T_4+T_5}{T_s} = -M\sin(\theta-\frac{\pi}{3})$ | 0 | $\frac{T_5}{T_s} = -M\sin(\theta)$ |

$$R_{(m+1)} = \frac{k}{f_x}f_s + (R_{(m)} + D_{(m)} - 1) \quad (15)$$

SNS-RP-SVPWM requires less memory occupation of the MCU than that in [13], since only the pulse position of the previous cycle must be stored, whereas the duty cycle of the previous cycle needs not be stored, as described in (11).

Fig. 3 shows the base space voltage vector diagram. The duty cycles of the three-phase switching pulse driven by 5-segment SVPWM is shown in Tables III, where $\theta$ is the phase angle of the output voltage vector, $M$ is the modulation index, $T_4, T_6, T_2, T_3, T_1, T_5, T_0$, and $T_7$ are the durations of the base space voltage vector in one cycle.

The amplitude of the output voltage vector will reach a maximum value equal to DC voltage link $u_{dc}$ if the motor speed increases continuously [14]. The modulation index is relatively close to 1 at a high rotation speed, resulting in the duty cycle of the three-phase switching pulse in Table III being close to 1 at certain positions, leaving less room for the selection of the pulse position. (14) shows that if $D_{(m+1)}$ is close to 1, $k_{\min} = k_{\max} + 1$ leads to no optional integer $k$ for calculating the pulse position of the subsequent cycle. Therefore, effective elimination of noise harmonic at a specific frequency may be infeasible under high rotation speed conditions in this method.

From (15), we can obtain:

$$f_x = \frac{k}{R_{(m+1)} - R_{(m)} - D_{(m)} + 1} f_s \geq \frac{1}{2 - D_{\min}} f_s \quad (16)$$

From (11), we can then obtain:

$$f_x = \frac{k}{R_{(m+1)} - R_{(m)} + D_{(m+1)} + 1} f_s \geq \frac{1}{2 + D_{\max}} f_s \quad (17)$$

It can be concluded that SNS-RP-SVPWM has a lower minimum limit for a specified frequency than that proposed in [13]. Meanwhile, as the switching frequency is constant, this method can quantitatively calculate the switching loss [15].

*B. SNS method based on a combination of RF-SVPWM and RP-SVPWM (SNS-RF-RP-SVPWM)*

Even with both the pulse position and switching frequency randomized, the correlation still existed in Methods 2 and 3, as shown in Table II. For example, $n = 1$ in Method 2:

$$f_{s(m+1)} = \frac{R_{(m+1)} + D_{(m+1)}}{\dfrac{k}{f_x} + \dfrac{R_{(m)}}{f_{s(m)}} - \dfrac{1}{f_{s(m)}}} \quad (18)$$

The value of the pulse position $R_{(m+1)}$ is selected randomly from the range represented by (12) with equal probability. In practical applications, the switching frequency needs to be limited to a certain range as follows:

$$f_{s\min} \leq f_s \leq f_{s\max} \quad (19)$$

By substituting (18) into (19), the range of $k$ can be written as:

$$\begin{cases} k \leq f_x \left( \dfrac{R_{(m+1)} + D_{(m+1)}}{f_{s\min}} - \dfrac{R_{(m)}}{f_{s(m)}} + \dfrac{1}{f_{s(m)}} \right) \\ k \geq f_x \left( \dfrac{R_{(m+1)} + D_{(m+1)}}{f_{s\max}} - \dfrac{R_{(m)}}{f_{s(m)}} + \dfrac{1}{f_{s(m)}} \right) \end{cases} \quad (20)$$

Similar to (14), while considering $k$ as a positive integer, we obtain:

$$\begin{aligned} k_{\max} &= \left\lfloor f_x \left( \dfrac{R_{(m+1)} + D_{(m+1)}}{f_{s\min}} - \dfrac{R_{(m)}}{f_{s(m)}} + \dfrac{1}{f_{s(m)}} \right) \right\rfloor \\ k_{\min} &= \left\lceil f_x \left( \dfrac{R_{(m+1)} + D_{(m+1)}}{f_{s\max}} - \dfrac{R_{(m)}}{f_{s(m)}} + \dfrac{1}{f_{s(m)}} \right) \right\rceil \end{aligned} \quad (21)$$

By substituting $k$ and $R_{(m+1)}$ into (18), we can obtain the switching frequency of the $(m+1)^{\text{th}}$ switching cycle.

Further, (18) is equivalent to:

$$R_{(m+1)} = \frac{k}{f_x} f_{s(m+1)} + \frac{f_{s(m+1)}}{f_{s(m)}} R_{(m)} - D_{(m+1)} - \frac{f_{s(m+1)}}{f_{s(m)}} \quad (22)$$

(22) is equivalent to (11) when the switching frequency is constant ($f_{s(m)} = f_s$). The value of the switching frequency $f_{s(m+1)}$ is selected randomly from the range represented by (19) with equal probability, and by substituting (22) into (12), the range of $k$ is as follows:

$$\begin{cases} k \leq f_x \left( \dfrac{1}{f_{s(m+1)}} - \dfrac{R_{(m)}}{f_{s(m)}} + \dfrac{1}{f_{s(m)}} \right) \\ k \geq f_x \left( \dfrac{D_{(m+1)}}{f_{s(m+1)}} - \dfrac{R_{(m)}}{f_{s(m)}} + \dfrac{1}{f_{s(m)}} \right) \end{cases} \quad (23)$$

Thus,

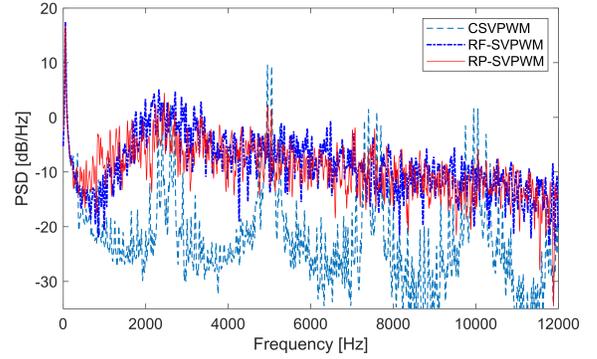

(a)

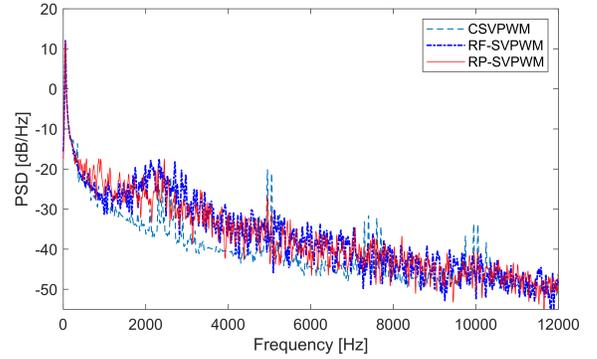

(b)

Fig. 4. Comparison of PSD using CSVPWM, RF-SVPWM, and RP-SVPWM. (a) Line voltage ($u_{AB}$). (b) Phase current ($i_A$).

$$\begin{aligned} k_{\max} &= \left\lfloor f_x \left( \dfrac{1}{f_{s(m+1)}} - \dfrac{R_{(m)}}{f_{s(m)}} + \dfrac{1}{f_{s(m)}} \right) \right\rfloor \\ k_{\min} &= \left\lceil f_x \left( \dfrac{D_{(m+1)}}{f_{s(m+1)}} - \dfrac{R_{(m)}}{f_{s(m)}} + \dfrac{1}{f_{s(m)}} \right) \right\rceil \end{aligned} \quad (24)$$

Similarly, the value of $k$ is selected randomly from $K = \{k_{\min}, k_{\min} + 1, \cdots, k_{\max}\}$ with equal probability. By substituting $k$ and $f_{s(m+1)}$ into (22), we can obtain the pulse position of the $(m+1)^{\text{th}}$ cycle.

There are two main advantages of (18) and (22). One is their ability to spread the power spectrum density (PSD) of voltage and current with better effectiveness owing to the use of variable pulse position and switching frequency. The other advantage is that, at a high rotation speed, (18) is also applicable and (22) is equivalent to RF-SVPWM.

## III. COMPUTER SIMULATION

The inverter shown in Fig. 1 driven by 5-segment SVPWM is simulated using MATLAB and Simulink in this section.

CSVPWM, RF-SVPWM, and RP-SVPWM methods from previous studies are simulated for comparison. In CSVPWM and RP-SVPWM techniques, the simulation parameters are as follows: the fixed switching frequency and DC voltage are equal to 2.5 kHz and 24V. The resistance and inductance of the loads are 1.02 Ω and 0.59 mH, respectively. The modulation index is 0.7 and the fundamental frequency of the line voltage is 50 Hz. The pulse position is located at the center of each switching cycle in CSVPWM, whereas in RP-SVPWM, it is randomly selected from 0 to $(1-D_{(m)})$. In RF-SVPWM, the switching frequency is selected randomly from 1.5 to 3.5 kHz with equal probability. Fig. 4 shows the PSD of the line voltage $u_{AB}$ and the phase current $i_A$. The harmonic noise is concentrated at the switching frequency and its integer multiples in CSVPWM, whereas in RF-SVPWM and RP-SVPWM, the voltage and current spectra are broadband, effectively suppressing the sideband harmonic noise.

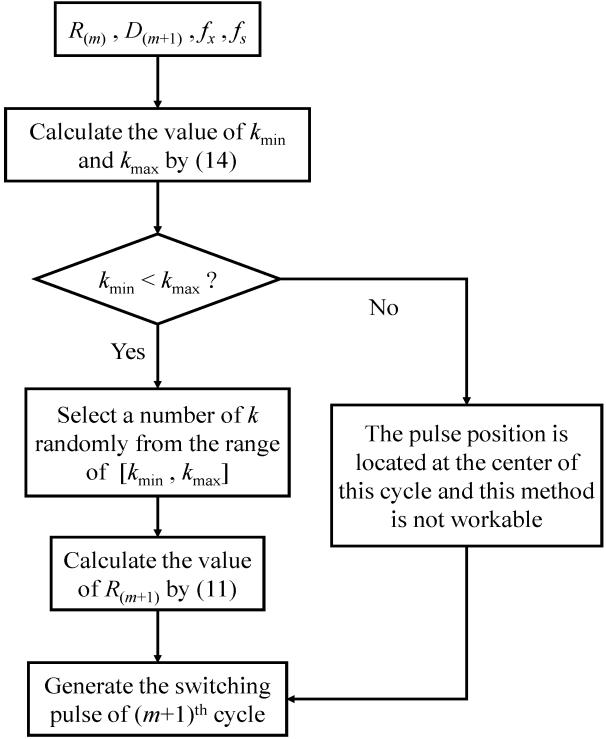

Fig. 5. Flowchart of the pulse position determination of SNS-RP-SVPWM.

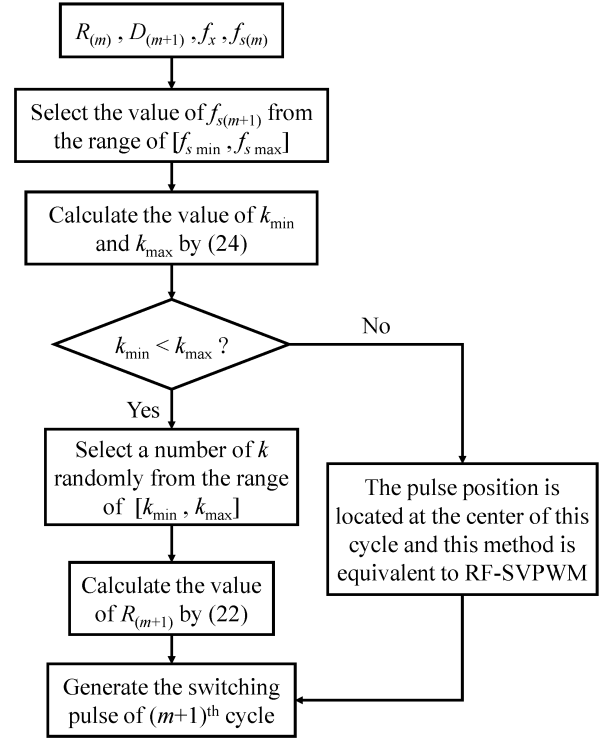

Fig. 7. Flowchart of the pulse position and switching frequency determination of SNS-RF-RP-SVPWM (using (22)).

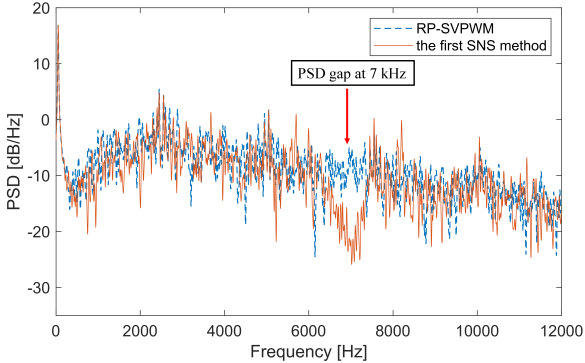

Fig. 6. PSD of the inverter line voltage ($u_{AB}$) using RP-SVPWM and SNS-RP-SVPWM ($M = 0.7$).

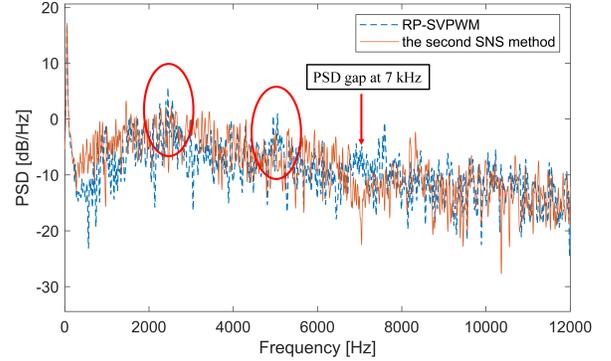

Fig. 8. PSD of the inverter line voltage ($u_{AB}$) using RP-SVPWM and SNS-RF-RP-SVPWM ($M = 0.7$).

*A. Verification and analysis of SNS-RP-SVPWM*

SNS-RP-SVPWM is explored as described herein. Flowchart of the pulse position determination for this method is shown in Fig. 5.

The desired specific frequency $f_x$ for elimination is equal to 7 kHz [16]–[18]. The remaining system parameters are similar to those of CSVPWM. The pulse position can be obtained using (11). The PSD of the inverter line voltage is shown in Fig. 6. Compared with RP-SVPWM, a gap at 7 kHz can be found in SNS-RP-SVPWM. The harmonic elimination band exceeds 1 kHz, so it is sufficiently wide to account for potential errors in the measurement and calculation of the resonant frequency. The maximum noise reduction at the specific frequency is approximately 15 dB/Hz.

*B. Verification and analysis of SNS-RF-RP-SVPWM*

Next, the effect of SNS-RF-RP-SVPWM on the voltage harmonics is explored. The procedure for determining the pulse position and switching frequency is illustrated in Fig. 7.

TABLE IV
MAIN PARAMETERS OF THE SPMSM

| Parameter | Value |
| --- | --- |
| Rated Voltage | 24 V |
| Rated Current | 4 A |
| Rated Speed | 3000 rpm |
| Rated Torque | 0.2 N·m |
| Rated Power | 62 W |
| Resistance | 1.02 Ω |
| Inductance | 0.59 mH |
| B-Emf Constant | 4.3 V/krpm |
| Pole number | 4 |

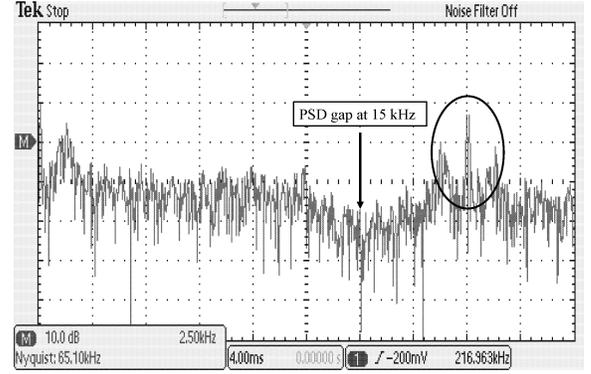

Fig. 10. Experimental results for PSD of the inverter line voltage using SNS-RP-SVPWM.

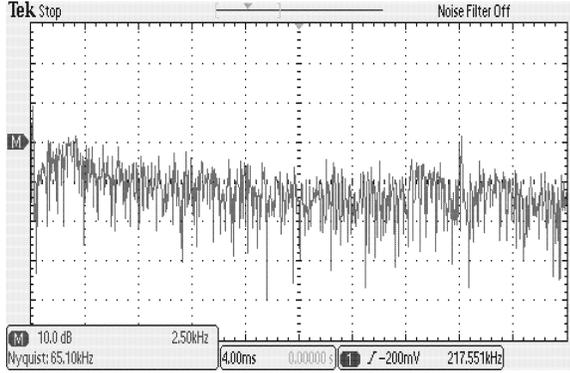

Fig. 9. Experimental results for PSD of the inverter line voltage using RP-SVPWM.

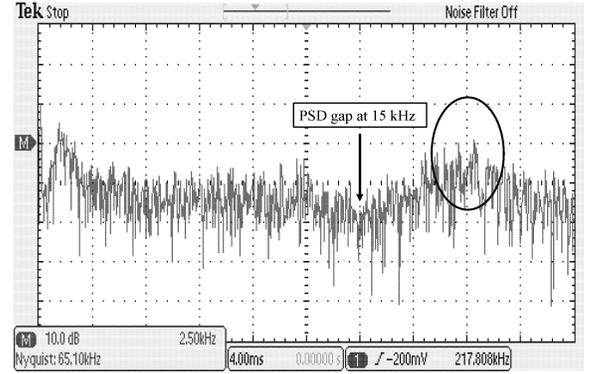

Fig. 11. Experimental results for PSD of the inverter line voltage using SNS-RF-RP-SVPWM.

The switching frequency is first selected from 1.5 to 3.5 kHz with equal probability in Fig. 7. The pulse position is then calculated according to (22). The specific frequency $f_x$ remains 7 kHz. The PSD of the inverter line voltage is shown in Fig. 8. Harmonic elimination at 7 kHz can be observed, with the maximum noise reduction being approximately 12 dB/Hz. Furthermore, the PSD using SNS-RF-RP-SVPWM is flatter near 2.5, 5, and 7.5 kHz compared to SNS-RP-SVPWM and RP-SVPWM. It should be noted that the overall voltage harmonics, especially at low frequencies, will increase.

## IV. EXPERIMENTAL RESULTS

The line voltage, vibration acceleration, and radiated noise spectra of a surface permanent magnet synchronous motor (SPMSM) driven by RP-SVPWM and the two proposed SNS methods are obtained and compared in order to verify the effectiveness of shaping the spectrum. The experimental setups are as follows: the three-phase inverter bridge circuit, using an Infineon N-channel power MOSFET IRFS3607, is used to drive the SPMSM. The main parameters of the SPMSM are listed in Table IV. An STM32F407 32-bit ARM MCU with a 168 MHz clock frequency is used to drive the inverter and the dead time is 100 ns. A Tektronix DPO2024 oscilloscope is utilized for the line voltage spectrum observation (X-axis: 2.5 kHz/Div; Y-axis: 10 dB/Div). A LAN-XI data acquisition module 3050-B-060 of Brüel & Kjær is used to measure vibration acceleration and acoustic noise in a semi-anechoic chamber. The type numbers of accelerometer and microphone are 352C22 and 4189A21, respectively. The sampling frequency is 51.2 kHz.

Fig. 9 illustrates the line voltage spectrum using RP-SVPWM when the switching frequency is set to 10 kHz, and the rotation speed of the SPMSM is 750 rpm. As previously noted, the PSD of the line voltage with RP-SVPWM is almost flat. The aim being to suppress the noise harmonics at a frequency of 15 kHz, experimental results confirm that the two SNS methods perform well. In SNS-RP-SVPWM, the switching frequency is constant at 10 kHz. The pulse position is then calculated according to (11). Contrastingly, in SNS-RF-RP-SVPWM, the switching frequency is selected in the range of 9 to 11 kHz, and the pulse position is calculated from the flowchart shown in Fig. 7. As shown in Figs. 10-15, a gap in the voltage, vibration acceleration, and radiated noise spectra at the desired frequency (15 kHz) can be clearly observed using both methods. The PSD in Figs. 11, 14, and 15 is more uniform and flatter near 20 kHz than that in Figs. 10, 12, and 13 obtained using SNS-RP-SVPWM. Therefore, SNS-RF-RP-SVPWM can spread the noise spectrum more effectively.

## V. CONCLUSION

The sideband noise of the inverter-powered motors obtained using CSVPWM technique is piercing, and random SVPWM methods for spreading the sideband noise can excite system

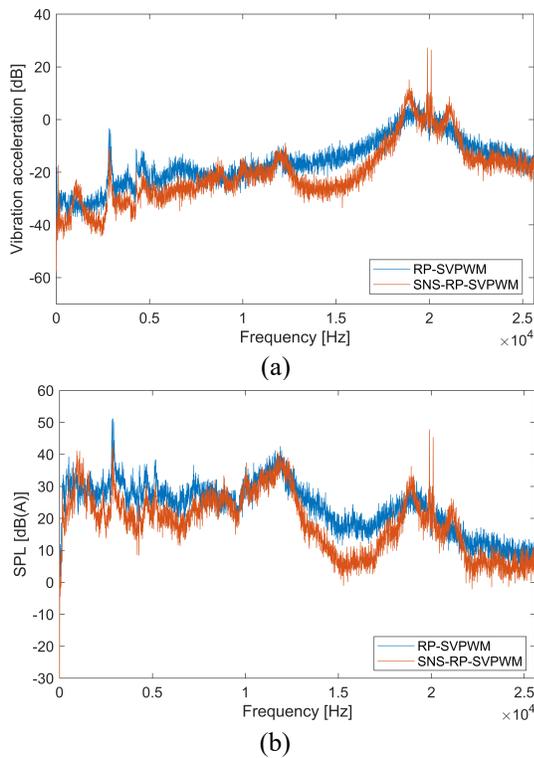

(a)

(b)

Fig. 12. The spectrum of the SPMSM annular housing with RP-SVPWM and SNS-RP-SVPWM at 750 rpm. (a). Vibration acceleration. (b). Noise.

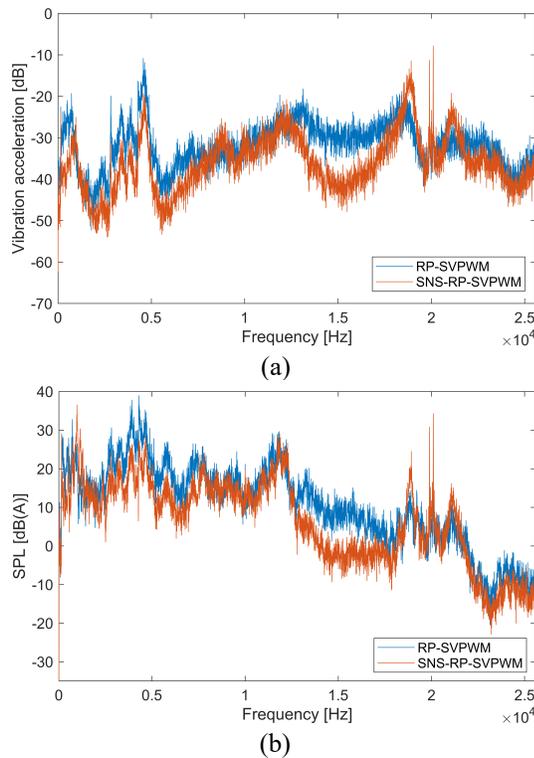

(a)

(b)

Fig. 13. The spectrum of the SPMSM disk-shaped cover with RP-SVPWM and SNS-RP-SVPWM at 750 rpm. (a). Vibration acceleration. (b). Noise.

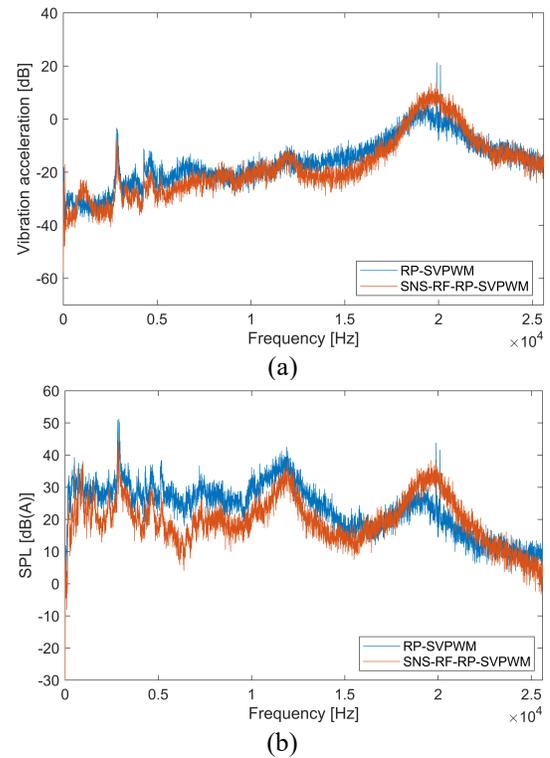

(a)

(b)

Fig. 14. The spectrum of the SPMSM annular housing with RP-SVPWM and SNS-RF-RP-SVPWM at 750 rpm. (a). Vibration acceleration. (b). Noise.

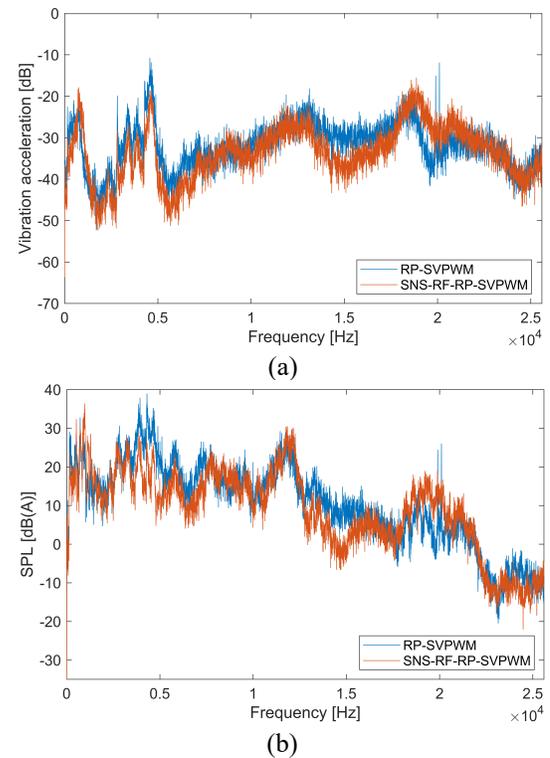

(a)

(b)

Fig. 15. The spectrum of the SPMSM disk-shaped cover with RP-SVPWM and SNS-RF-RP-SVPWM at 750 rpm. (a). Vibration acceleration. (b). Noise.

resonance at modal frequencies. To solve these problems, this study derives the general formula of SNS methods and proposes two SNS methods to eliminate noise harmonics. SNS-RP-SVPWM based on RP-SVPWM shapes the voltage spectrum successfully. It requires less memory occupation of the MCU and has a lower minimum limit for a specified frequency than that proposed in [13]. Nevertheless, similar to RP-SVPWM and the method proposed in [13], it is only effective under the low rotation speed conditions, and noise harmonic at the switching frequency is significant. SNS-RF-RP-SVPWM is based on a combination of RF-SVPWM and RP-SVPWM, and establishes a unique correlation among the switching frequency, pulse position, and duty cycle to suppress voltage harmonics. It spreads the PSD of voltage with better effectiveness. However, it can only estimate the switching loss because of random switching frequency. Computer simulation and experimental results prove that both proposed SNS methods can effectively perform selective noise suppression at a specific frequency. Notably, the overall voltage harmonics, especially at low frequencies, are expected to increase. These SNS methods serve as useful references to prevent the excitation of the resonant frequencies of motors, and provide theoretical and technical guidance for future applications of low-noise motors.